\documentclass{article}

\usepackage[safe]{silence}
\usepackage{arxiv}

\renewcommand{\undertitle}{}
\renewcommand{\headeright}{}
\date{} 
\renewcommand{\thefootnote}{\arabic{footnote}} 
\renewcommand{\fnsymbol}[1]{\arabic{#1}} 

\usepackage[utf8]{inputenc}
\usepackage[english]{babel}
\usepackage{url}
\usepackage{xurl}
\usepackage{booktabs}
\usepackage{amsfonts}
\usepackage{amsmath}
\usepackage{nicefrac}
\usepackage{microtype}
\usepackage{xcolor}
\usepackage{graphicx}
\usepackage{makecell}
\usepackage{threeparttable}
\usepackage{multirow}
\usepackage{float}
\usepackage{longtable}
\usepackage[round]{natbib}
\usepackage[breaklinks=true]{hyperref}
\usepackage{draftwatermark}

\begin{document}
\title{There to care; not to kill: medical settings, statistics and wrongful convictions}

\author{
Prof.dr. Richard D. Gill\thanks{Preprint. Not yet reviewed or accepted for publication} \\
Mathematical Institute, Faculty of Science \\
Leiden University \\
\texttt{gill@math.leidenuniv.nl}
}
\date{2 March, 2026}

\maketitle
\renewcommand{\thefootnote}{\arabic{footnote}}

\begin{abstract}
This paper discusses wrongful convictions in a medical setting, focusing on nurses. Common features are lack of strong direct evidence: the nurse was never seen doing anything wrong. There is no DNA evidence of tampering of apparatus or medications by the nurse. There is no CCTV footage showing suspicious actions. Analysis of medical records at the time led coroners to issue certificates of natural deaths, and most events were not, at the time, thought suspicious by hospital staff. There is no confession and the nurse consistently asserts they are completely innocent. There is no evidence of earlier psychopathic behaviour. Instead, private writings (e.g., in a diary) are interpreted by the prosecution as a confession; mundane behaviour is given a sinister interpretation. Motive remains speculation. The main evidence is statistical: a spike in deaths or collapses and a statistical association with a particular nurse. There is forensic evidence which suggests one or two patients might have been harmed by administration of medication much used in the hospital, and even legitimately used earlier in the care of the alleged victims. Police investigations are driven by the hospital consultants who were clinically responsible for the patients allegedly killed or harmed by the nurse.
\end{abstract}

\section{Introduction}

Consider a hospital ward where patients are cared for whose death is a
not unlikely outcome: for instance, a ward occupied by the very aged and
infirm, the terminally ill; or a neonatal intensive care unit where
extremely premature babies are hopefully nursed to ``full term'' and
hospital discharge, but where inevitably some are going to succumb to
fatal complications and die. All the time, external circumstances may be
changing in known or unknown ways; for instance, the incoming stream of
new patients might change in character due to changes in capacity in
neighbouring hospitals. Internal organisational changes are also
possible. Moreover, a ward can be struck by some initially undiagnosed
infection leading to an increased death rate of vulnerable patients. In
summary, an apparently stable and low death rate on a medical unit can
suddenly increase for no immediately obvious reason, producing what
appears to be a suspicious cluster of bad outcomes.

The exact hour a seriously ill person is going to die is usually quite
unpredictable (and similarly, for collapses). The exact cause of death
of a seriously ill person usually cannot be determined; there is
typically a cascade of deterioration of many of the organs and systems
of the body. Thus, deaths can be unexpected and initially unexplained.
Death certificates contain global descriptions such as ``frailty of old
age'' (in the case of the author's mother at 97 years old) or ``extreme
prematurity'' (in the case of some of the babies whom neonatal intensive
care nurse Lucy Letby in the UK is supposed to have murdered without
anyone have seen her done anything wrong).

A hospital is a place with a strongly hierarchical workforce. Highly
educated and highly paid consultants are responsible at a high level for
clinical management of the whole ward as well as being involved in the
care of individual patients. Most day-to-day care of patients is done by
poorly paid nurses. In between there are usually many young, trainee
doctors. Mistakes are easy to make, hard to admit.

Medical diagnosis and corresponding course of treatment is not an exact
science, but a ``best guess''. It has to be based, in real time, on
incomplete and imperfect information. Unexpected side effects or other
unexpectedly arising complications might show that the best guess
treatment is actually causing the patient harm, and possibly even death.
Patients suffering complex diseases will be cared for by a
multidisciplinary team of medical experts and not all have access to all
the medical information about the patient, especially when patients are
transferred from one hospital to another.

There tends to be a culture of denial concerning medical errors. Senior
doctors never criticise other senior doctors. Concerning some unexpected
or disturbing event, each one will tell the truth about what they know
about it, but sometimes not the whole truth; they will avoid suggestions
that a colleague might be to blame. Errors are frequently attributed to
nurses. Nurses do not exhibit strong collegiality, many can be
indifferent to a colleague getting the blame, and ``help'' consultants
or police investigators by telling the truth but not the whole truth.
When police investigations start, especially the nurses will be coached
by hospital lawyers.

All this can easily lead to the occurrence of disturbing sequences of
deaths or collapses which at the time are hard to explain. People see
patterns. Once one rare event is seen, people will start seeing it
repeated due to a well-known cognitive bias: the Frequency illusion or
Baader-Meinhof phenomenon \citep{wikipedia2026}.

Hospitals are very dangerous places, see \citet{zaslow2023}. According to the
\citet{worldhealthorganisation2023}\begin{quote}

Around 1 in every 10 patients is harmed in health care and more than 3
million deaths occur annually due to unsafe care. In low-to-middle
income countries, as many as 4 in 100 people die from unsafe care.\end{quote}

Above 50\% of harm (1 in every 20 patients) is preventable; half of this
harm is attributed to medications.

Some estimates suggest that as many as 4 in 10 patients are harmed in
primary and ambulatory settings, while up to 80\% (23.6--85\%) of this
harm can be avoided.

Common adverse events that may result in avoidable patient harm are
medication errors, unsafe surgical procedures, health care-associated
infections, diagnostic errors, patient falls, pressure ulcers, patient
misidentification, unsafe blood transfusion and venous thromboembolism.

Though the WHO report mentions ``low to middle income countries'' one
may suspect that the statistics in high income countries are not much
different, especially in high income countries with large income
inequality (the US, UK). In the UK it is widely agreed that the nation's
pride, the NHS (National Health Service), is underfunded and
incompetently run. Failing maternity care is currently a focus of UK
government attention. It has been argued that 1000 deaths per year in
maternity care could be avoided in the UK, were care of the same
standard as in Sweden. On the other hand, much of the huge difference in
perinatal mortality and mortality of mothers associated with pregnancy
and birth is explained by socio-economic factors, \citet{zylbersztejn2018}.

A cluster of (at that time) unexpected and (at that time) unexplained
events will typically not be enough to trigger a police investigation.
However, hospital authorities may well immediately have the possibility
of a serial killer in their minds (in the UK thanks to the cases of
family doctor Harold Shipman and nurse Beverley Allitt). The first step
will be an internal hospital investigation into commonalities between
those events which might help explain them. Looking at nursing rosters
will likely be part of that. Senior medical staff (the specialists /
consultants) will be tasked with studying the medical notes of the
patients in their care. This can be the stage at which the idea that a
particular nurse is linked to the cluster may first arise; and that
could well be a nurse who is particularly noticeable due to character,
behaviour \ldots{} and frequent presence due to long working hours. At a
2017 meeting of the management team of the Countess of Chester Hospital
(CoCH) and the hospital trust board to discuss whether to call in police
and if so, what to say to them, the chairman of the board explicitly
brought up the name of Beverley Allitt. The recent spike in deaths and
unexplained collapses in the neonatal intensive care unit (NICU) was
already explicitly associated with nurse Lucy Letby through inspection
of nursing rosters, though initially another nurse seemed also to be
present just as often at the same events, see the CoCH trust and
hospital management documents and email conversations between
consultants and managers \citep{thirlwallinquiry2026}.

Where people are dying and mistakes are common, it can be very difficult
to distinguish malicious harm, accidental harm, and unavoidable
``adverse events'' such as death. On the one hand, this would provide an
evil person with opportunity: opportunity to kill or otherwise cause
harm while avoiding detection. It has been argued by criminologists
studying health care serial killers (HCSKs) that this is the reason for
the relatively high proportion of HCSKs in prison populations. For
instance, the proportion of serial killers who are nurses in UK prisons
is 10 times the proportion of working nurses in the population (though a
large part of this over-representation is due to length bias: most
people going to jail in any period are not serial murderers; but once in
prison the serial killers stay there the longest). According to
criminologists and forensic psychologists, many of such crimes are
committed by a person suffering from the controversial psychiatric
diagnosis ``von Munchausen by proxy''. Forensic psychologists satisfy
the public hunger to understand the mind of horrific killers. They
consult with police investigators in profiling, in the stage at which
there is a clear cluster of murders but no suspect yet identified, or
they later advise a court.

The phenomenon of serial killer nurses (usually female) is of enormous
fascination to the public. This has engendered a whole research field
and the leading researchers in this field become go-to experts for the
media. Criminologists, by definition, study criminals, and naturally
gather data on convicted criminals, irrespective of whether they are
justly convicted or not; criminals who were not detected or who were
only briefly ``persons of interest'' do not enter into their data bases.
Research is mainly based on newspaper reports and judicial publications
following successful convictions which naturally focus on the
prosecution's picture. Two key publications are \citet{yorker2006} and
\citet{yardley2014}, which catalogue past cases and look for
statistically recurring features. Embarrassingly, these papers also
relied on convictions which were later overturned and others which are
still highly contested, for instance the case of Lucia de Berk in the
Netherlands, to which we will return. The resulting ``red flag lists''
which are supposed to help hospital managers and doctors to spot a
killer nurse before he or she has harmed very many patients include the
very features used by the prosecution as evidence against Lucia de Berk.
The red flag lists contain therefore the features which led to
``successful prosecution''. They hereby help us see how wrongful
convictions arise, by identifying the kind of aberrant or striking
behaviour of a nurse which can lead them to be associated with a spike
in deaths or other adverse events on a hospital ward.

Serial killer nurse cases almost always involve statistics, whether
explicitly or not. The impetus is a cluster of bad outcomes, and
``cluster'' is a statistical concept. Once attention has been placed on
nurses, the administrative data of which nurses were on shift and when,
will come into play. Investigators will look for an association or a
correlation between events of interest, and presence or absence of
nurses. Even if those investigators are supported by professional
statisticians or epidemiologists, they can easily fall prey to mistaking
correlation for causation. If this mistake is made at an early stage of
investigations, the results may be confirmation bias and the Texas
sharpshooter paradox.

In short: there is a spike in (say) deaths. They appear to be associated
with a particular nurse. Events (not just those in the initial
``cluster'') are re-evaluated with the role of that particular nurse in
mind. If he or she is there, then reasons are seen to find the event
suspicious (unexpected and unexplained). Events when she is absent are
not catalogued; not even searched for.

The author of this chapter has been formally involved as a defence
expert witness in applications to the CCRC for the UK case of Ben Geen.
He has some familiarity with the UK cases of Victorino Chua and of Colin
Norris and is currently highly focussed on the case of Lucy Letby, as a
campaigner for a retrial. He played a major role in the cases of Lucia
de Berk (Netherlands) and Daniela Poggiali (Italy). The latter were both
exonerated after many years in jail.

At the time of writing, the highly controversial Lucy Letby case has not
been legally determined to have been a wrongful conviction. Ms Letby is
currently in prison with a total of 14 full-life sentences, having been
denied an appeal after either of two trials. A new defence team has
applied to the CCRC to have the case reopened; it contains materials
prepared by more than 20 experts in numerous fields. Those experts
presently avoid speaking in public. Those presently expressing concern
in public include a former Supreme Court judge, senior barristers and
solicitors, two former government ministers, numerous eminent professors
of medicine and statistics, and high-ranking (former) police officers.

An important feature of all these cases is: nobody ever saw the nurse in
question do anything wrong. All the nurses staunchly continue to declare
their innocence.

It is useful to contrast these cases with the cases of two convicted
health care serial killers whose convictions are either not disputed at
all, Charles Cullen (US), or hardly disputed, Beverley Allitt (UK).

How can a jury or, in an inquisitorial system, a board of wise judges,
find a person guilty of premeditated murder when there is no direct
evidence of their ever having harmed a patient in any way at all? When
all deaths, at the time they happened, were found by coroners to have
been natural deaths?

There is necessarily a mountain of circumstantial evidence, in
particular: the previously mentioned spike in unexplained deaths, and
the correlation of bad events with the presence of the bad nurse.

There is a mountain of subjective interpretation of behaviour. For
instance, Lucy Letby was accused by the prosecution barrister of crying
on the stand when cross-examined on the loss of her nursing career, but
not when cross-examined on the babies who had died while in her care.
This was painted by the prosecution as evidence of her evil nature. Her
behaviour in court was withdrawn, ``colourless''. Commentators wrote
that an innocent person would surely have screamed out their innocence
and knocked over chairs and tables during police interviews, and later
the stand. Lucy's demeanour at work -- careful and meticulous -- was
interpreted as being a deliberate ruse to allow her to commit her crimes
unnoticed. Her eagerness to do overtime and eagerness to care for the
most seriously ill infants was similarly given a sinister
interpretation. No account was taken of the fact that by the time she
appeared in court she was suffering from PTSD, medicated, and had been
persecuted continuously by powerful persons and organisations for a full
seven years.

In the disputed cases, a commonality is the complete lack of evidence of
psychiatric disorder. Even the prosecution agrees that they do not
understand the reason why the convicted person carried out those evil
crimes. On the other hand, already in his childhood, Charles Cullen
carried out sadistic attacks on small animals. As a teenager, Beverley
Allitt practised self-harm. Cullen admitted to his crimes in a taped
conversation with his girlfriend. Allitt's confession is not reliable
evidence since it secured her a more friendly prison confinement in a
closed psychiatric prison where she would be safer -- as a woman who was
convicted of killing children -- than in a regular prison. Allitt did
not reveal how she had harmed various patients - her confession
contained nothing which was not already public knowledge.

The prosecution case is typically almost entirely based on
circumstantial evidence and subjective evaluation of behavioural
patterns. However, there is typically in these disputed cases a
``smoking gun'' of forensic evidence, which convinces the court that one
or more deaths were murders. It remains to determine the perpetrator. In
the Letby case there are two babies for whom an immunoassay suggested
poisoning with insulin (actually, there was a third such case on the
unit, but Letby was not charged with harming that child!). In the case
of Lucia de Berk there is one child for whom a post-mortem forensic test
suggested digoxin poisoning. In the case of Daniela Poggiali there was a
post-mortem forensic test suggesting potassium chloride poisoning.

\section{Forensic statistics, and the RSS report} Statisticians and epidemiologists are frequently called as expert witnesses in serial killer nurse cases, even though the legal community finds them particularly hard to understand. The Lucy Letby case concerns events from June 2015 to June 2016. We will say a few words on the genesis of the case later. Statisticians are often asked to analyse data which has already been collected by other parties, and the first task is to ascertain how and why that data was gathered. A brain teaser like the three-door problem (the Monty Hall problem) exemplifies that data cannot necessarily be taken at face value. How it came to be selected or generated can dramatically influence how it should be interpreted. In the Letby case, a police investigation started in 2017, and it was immediately reported in the newspapers. Very early on, the key features of an inexplicable spike in deaths together with an association with a nurse, was public knowledge.

At conference and workshop meetings of forensic statisticians the case
was mentioned in informal discussions. The misuse of statistics in the
cases of Lucia de Berk, Daniela Poggiali and Ben Geen was already well
known, see \citet{meester2007}, \citet{gill2018},
\citet{dotto2021}, \citet{gill2022}.

The UK's Royal Statistical Society is one of the most prestigious
statistical societies in the world, devoted to interaction between
theory and practice. The UK Sally Clarke case had earlier led to the
founding of an RSS section on Statistics and the Law. As the Letby case
started regularly featuring in news reports, the then current
chairperson Prof. Jane Hutton set up a working group of five, chaired by
Prof. Peter Green (who had been president of the RSS during the time of
involvement with the Sally Clark case), and with members Prof. William
Thompson (a US lawyer), Prof Julia Mortera (Italy, a statistician),
Prof. Richard Gill (Netherlands, a statistician), and Mr. Neil Mackenzie
KC (a Scottish barrister). Notice the mix of jurisdictions and
expertise. The report \citet{royalstatisticalsociety2022} was published
months before the trial of Lucy Letby started, and without any idea of
what evidence was going to be brought into the trial, without any
information on how statistical data was collected by the police, nor
whether or how it was going to be used.

The authors' aim was to pinpoint the danger of confirmation bias and of
faulty interpretation of correlation especially when based on biased
data collection. It aimed to help all parties involved by giving
guidance on how to avoid the statistical pitfalls, which are actually
well-known pitfalls in research methodology in general. ``Data
collection'' should be seen broadly; it includes the definition of cases
of interest, which naturally must be done in medical terms, but as
statistical researchers in the field of medicine know, requires
collaboration between medical and statistical experts.

In a press release, the RSS stated ``\textit{The report calls for more care to be taken by experts to avoid drawing erroneous inferences from such data, by properly controlling for plausible causal factors.~The RSS's concerns about the compilation of the data used in such investigations are that attention is rarely given to ensuring that unconscious bias has not influenced the selection of cases.~Such innocent cognitive biases are prevalent throughout society and control of these needs active steps such as blinding.~For~medical misconduct~cases, the report recommends that~investigations should be supervised by expert panels independent of both the suspect and their employer.}''

The report includes several case studies. It makes eight recommendations
and gives extensive reasoning for each of them, as well as cautionary
tales concerning what has in the past happened when they were not
followed.

\textbf{Recommendation 1}: \textit{It is therefore important that all parties involved in investigation and prosecution in such cases consult with professional statisticians and use only such appropriately qualified individuals as expert witnesses}.

\textbf{Recommendation 2}: \textit{In presenting the results of statistical tests, both the level of statistical significance (p-value) and the estimated effect size should be stated. One addresses the question of whether an effect is truly detected},

\textbf{Recommendation 3}: \emph{In reports and testimony, experts
should take care to explain the proper interpretation of p-values and
should avoid drawing fallacious inferences from them. In jurisdictions
that rely on lay jurors, judges should consider providing instructions
about the proper use of p-values. Lawyers, judges and investigators
should educate themselves to the dangers of fallacious statistical
interpretation. Lawyers should endeavour to present the case in a manner
conducive to correct understanding, avoiding to the extent possible
testimony or arguments conducive to misinterpretations}.

\textbf{Recommendation 4}: \emph{Investigations should be guided by
panels representing all relevant areas of expertise but independent of
both the suspect and the employing institution.}

\textbf{Recommendation 5}: \emph{To the maximum extent practicable,
experts informing an investigation, such as DNA specialists, fingerprint
examiners, toxicologists, and pathologists should be kept  ``blind''
to all aspects of the case irrelevant to the question they are being
asked to answer. Blinding is a key tool in minimising prejudicial
subjective effects such as unconscious bias.}

\textbf{Recommendation 6}: \emph{It is vital that investigators
appreciate the truth of this, and the fact that the connection between
them is well-studied, and that in fields such as medical diagnosis there
are accepted criteria to guide the valid drawing of conclusions in
observational studies. Possible confounding factors must be identified,
and their effect quantified, before attributing causes to observed
effects.}

\textbf{Recommendation 7}: \emph{When courts must evaluate the results
of problematic investigations, it is particularly important that they
consider reports and expert testimony from independent statisticians. If
investigative bias is a significant concern, lawyers and courts should
also consider seeking evaluations from experts of cognitive bias and
factors associated with the accuracy of expert judgment.}

\textbf{Recommendation 8}: \emph{Further interaction between legal and
statistical communities should be fostered by the leaders of the legal
and statistical communities, with a view to promoting joint educational
activities.}

It is unfortunately by now clear that the language of the report went
totally over the heads of those involved in the Lucy Letby prosecution.
The prosecution \emph{deliberately} did not employ a professional
statistician and \emph{deliberately} did not conduct any formal
statistical analysis. Police investigators at one point did plan to
``instruct'' a prominent medical statistician (the afore-mentioned Jane
Hutton) but were told by the CPS (Crown Prosecution Service) to ``drop
that line of inquiry''. The police had in fact asked if the statistician
could give them something like ``a one in a ten million chance'' that
the correlation was a coincidence. It is unlikely to be a coincidence
that this is exactly what professor of medical statistics Carol Jagger
had done in the Beverley Allitt case. A senior member of the CPS told
this author that ``they are not using statistics, because it only
confuses people''. The police however have consistently referred to
their analysis of the relationship between events and Lucy Letby's
presence as a sophisticated statistical analysis, performed under
contract by a forensically accredited IT company.

It is now time to turn to examples, organised according to the triad of
``statistical evidence'', ``forensic evidence'', ``character evidence'',
and focussing on the items which convinced the public of guilt and
likely convinced court experts of guilt and the final fact-finders,
whether a jury in an adversarial system or a board of judges in an
inquisitorial system. In fact, since the judge plays a huge role in UK
courts in summarising the evidence presented to the jury and in
instructing the jury how they might think about it, it is the present
author's opinion that there is no major difference between the two
systems in practice. In both systems, during a trial, prosecutors play
to the public and are fully aware of the importance of public opinion,
even though the official position is that it plays no role whatsoever.
Police investigators and wise judges are just as susceptible to moral
outrage at the suggestion that a nurse is murdering helpless patients,
and susceptible to confirmation bias. Statistical evidence and
probabilistic reasoning is not understood by police, lawyers,
journalists, or the public. The Texas sharpshooter paradox played a role
in numerous convictions.

\section{Statistical evidence}

We start with an example, taken from a case which is presently not seen
as unsafe and which has instead delivered, in the UK at least, an
archetype female serial killer, and a playbook for future investigations
and trials. This is the case of nurse Beverley Allitt, subject of
several books and TV documentaries, many of them now hard to obtain or
view. However, \citet{ross2025} is recent and comprehensive. The roster chart
which follows was reconstructed by L.~Turnbull MA (March 2026, India) from screenshots
from two TV documentaries. It is available from the present author as a
.xlsx file.

\begin{figure}[t]\centering\includegraphics[width=\linewidth,height=0.4\textheight,keepaspectratio]{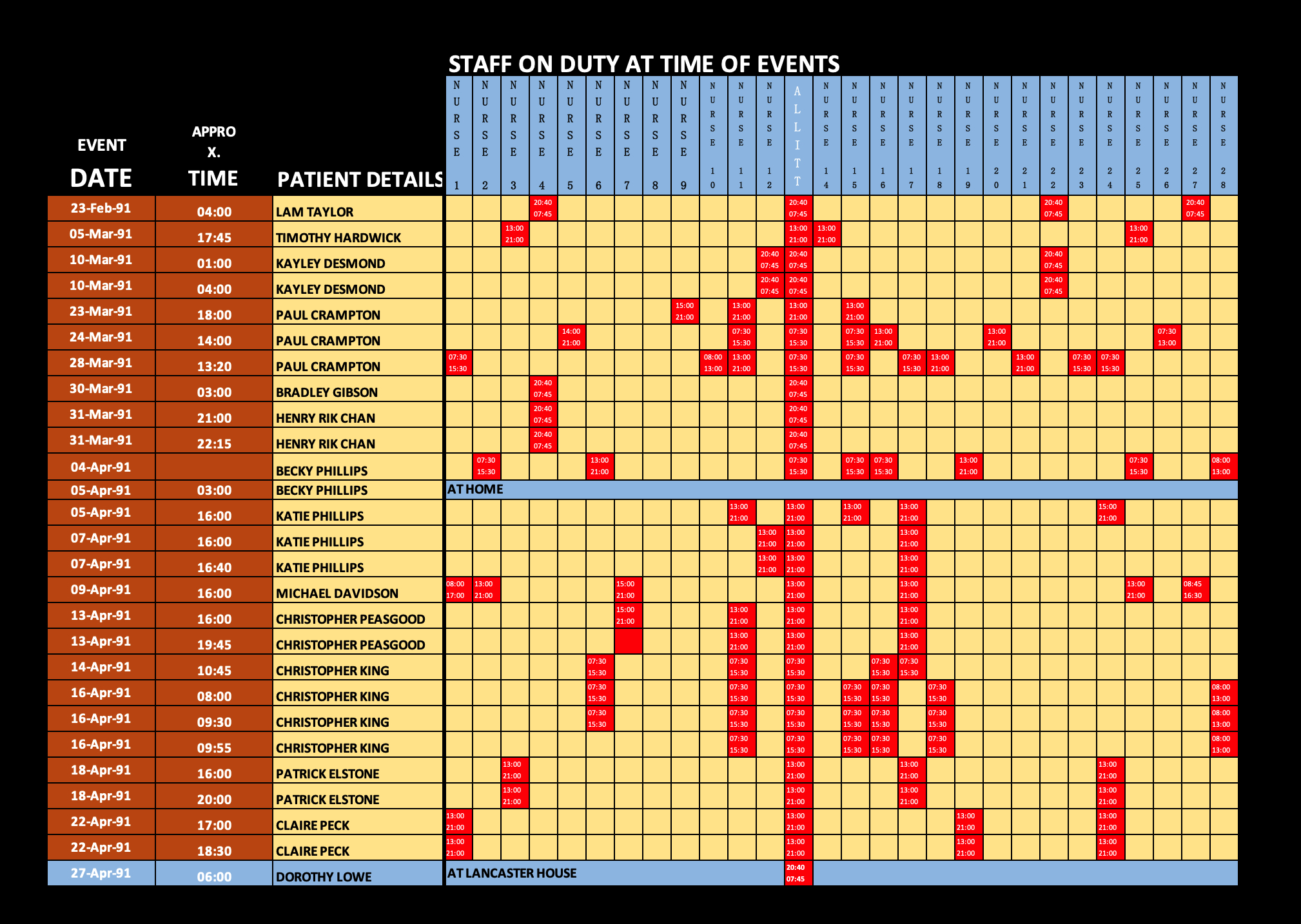}\caption{the Bev Allitt roster chart.}\end{figure}

Each column of the spreadsheet belongs to one nurse; each row belongs to
one indictment. Only one nurse is identified by name, nurse 13 (!). In
the TV documentaries all names (except for Allitt's) are blurred, but
the ordering is clearly not alphabetical. Allitt was charged with, and
found guilty of, four counts of murder, eleven counts of~attempted
murder, and eleven counts of causing~grievous bodily harm.

Here is a second example, the roster chart which played an important
role in the case of Lucy Letby.

\begin{figure}[t]\centering\includegraphics[width=\linewidth,height=0.4\textheight,keepaspectratio]{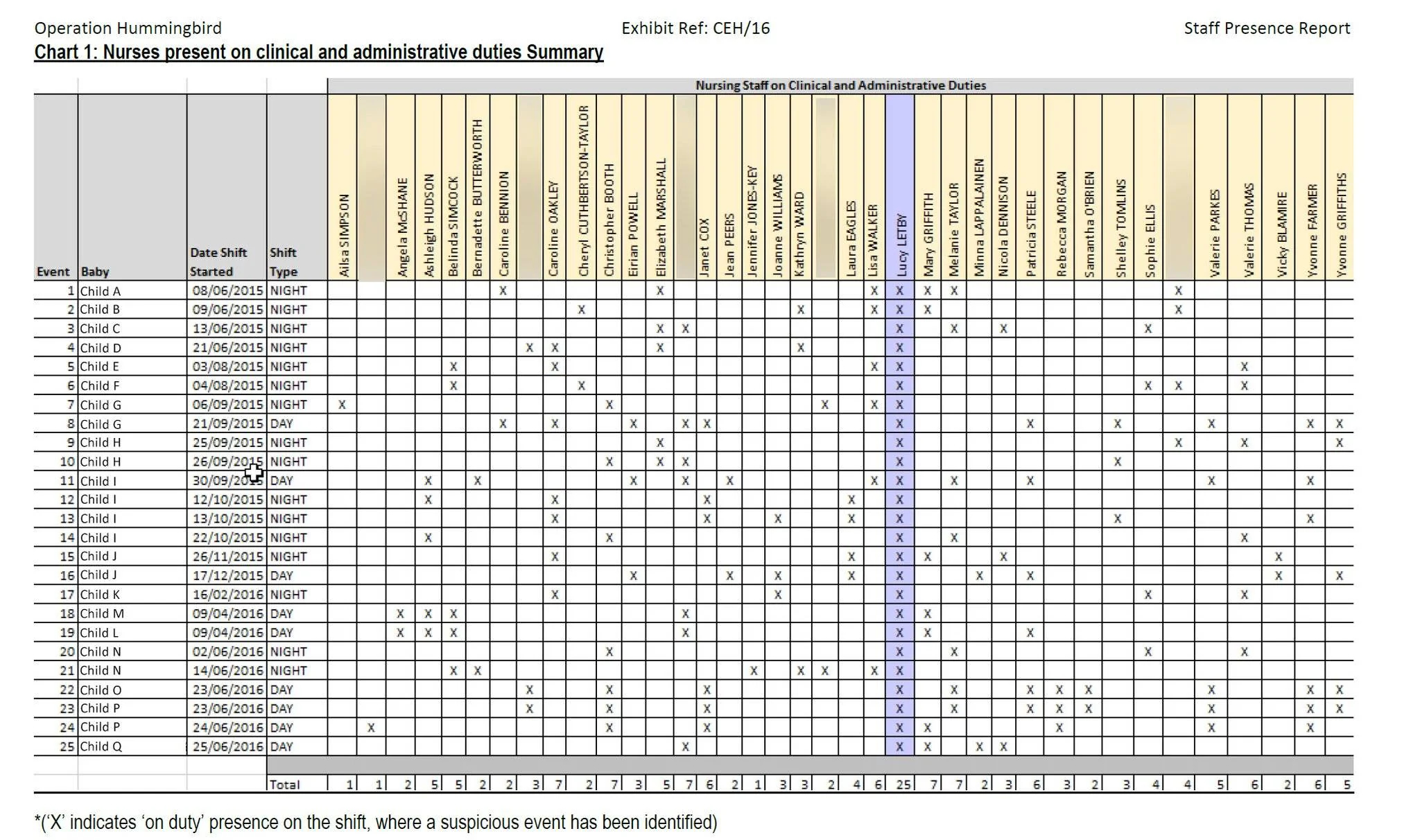}\caption{the Lucy Letby roster chart.}\end{figure}

Again, columns are nurses, rows are events. This time the nurses are in
alphabetical order, Letby is automatically nicely in the middle of the
table. The events again correspond to indictments on the charge sheet.
The jury returned a small number of ``undecided'' and of ``not guilty''
verdicts. After a second trial on one of the ``undecided'' cases, she
had finally received 14 full life sentences for 7 murders and 7
attempted murders.

The graphic was shown to the jury on the first day of the trial, and
again, on the last. Journalists asked permission to reproduce it in the
media, the judge permitted them to do so. The prosecution case was that
Lucy Letby was the common denominator in \emph{all} suspicious deaths
and collapses. Actually, early in the trial one of the events had to be
scrapped. Lucy was not present. One can find a corrected spreadsheet in
the files of the \citet{thirlwallinquiry2026}, with one less row.

Merely by the existence of those two versions, we know that the police
spreadsheet contains an important error: at least one event was
identified as suspicious, and Letby charged with causing that event
deliberately, even though she had not been present at all. Later
revelations showed that there was not just one ``missing'' suspicious
event: the prosecution's star expert witness, retired paediatrician Dr
Dewi Evans, had identified about 20 further suspicious events, for none
of which Letby was charged. Whether or not she was involved in those
babies' care is unknown.

The Allitt and the Letby spreadsheets or roster charts are a powerful
visual aid in getting across a statistical point to a jury and to the
public. Those two nurses are always around when bad things happen. They
are examples of the famous saying ``lies, damned lies, and statistics''.
They are strongly suggestive of things which are quite simply not true
in the Letby case, and possibly not true in the Allitt case either.

From a legal point of view, the charts are irrelevant. Each individual
indictment is tested in court, and allegedly each one is tested
``independently'' on its medical merits together with character evidence
obtained from all the cases.

A statistician will immediately ask how were suspicious events defined
and identified? In neither case is the answer fully known. In the Letby
case, her defence barrister asked for such a specification, but the
judge turned down the request. From the judge's point of view, here are
25 (or 24) indictments and each one simply must be evaluated separately
on the basis of the medical evidence pertaining to that event, and only
that event. The Thirlwall Inquiry published the shorter spreadsheet. The
terms of reference of the inquiry are that Letby is guilty and the fact
that she was not found guilty of many charges, and even that a charge
was dropped during the proceedings, is of no interest. The process by
which charges arrived is of no interest. Fortunately, however, the
inquiry has unearthed a treasure trove of information precisely
concerning that issue.

In the Allitt case a statistician, Prof. Carol Jagger, helped underscore
the visual impact of the chart by coming up with a powerful probability
statement ``less than one in 10 million''. The number is irrelevant (and
how it was computed is unknown). It is crystal clear that Allitt's
presence every time, standing out from all the other nurses, is not due
to chance. However, the computed probability has powerful impact on
everyone who hears it, including on medical and forensic experts. Prof
Vincent Marks (private communication) told this author that he was
personally convinced that Allitt was guilty because of the roster chart.
As a forensic toxicologist, he was convinced that certain patients had
suffered factitious insulin poisoning, that is, poisoning with insulin
which they were not supposed to have been given. He reports in the book
\citet{marks2007} that a pretrial expert meeting at which his
insulin and Jagger's statistical findings were both presented convinced
all 50 participants of Allitt's guilt (including police investigators
and the two paediatricians at the hospital).

In fact, both tables contain interesting but somewhat hidden statistical
information. In the Letby case, it simply is not the case that during
the period of interest, 38 very similar nurses were working roughly
similar hours and with roughly similar duties. Many of the 38 only work
part-time or are bank nurses (previously employed at the hospital,
prepared to come back in times of high activity), or agency nurses. Very
few are full time, fully engaged directly in patient care, throughout
the periods of interest. In the Letby case there could be five such
nurses including Letby herself. We know that only one of them was as
qualified as Letby. The most qualified nurses get allocated to the most
severely ill patients. Inspecting the numbers of suspicious events each
nurse experiences, one clearly sees two groups of nurses, five or so
with a large number, the remainder with a small number. \citet{oquigley2025}
has published a sophisticated statistical analysis which shows that,
taking account of the ``missing rows'' of the spreadsheet, the nurse
with \textit{by chance} the largest number of ``suspicious events'' in
their shifts would likely have about the same number as Lucy Letby had.
One of the five comparable nurses must have the biggest number! Once we
are looking for a killer nurse, analysis of roster data will find that
one, and provide us the damming roster chart, simply through the Texas
sharpshooter paradox.

So far, we have not mentioned numerous other problems with the charts.
In the Letby case there is a large preponderance of twins and one
triplet, together making up nearly half of the babies in the indictment.
One does not need to be a statistician to realise that if one of a pair
of identical twins experiences some collapse that their sibling could
well be equally vulnerable. We now know that the identical twins, and
the triplet, are monochorionic (share a placenta) and hence vulnerable
to some less known but extremely dangerous complications, such as TTTS
(twin to twin transfusion syndrome). Medical experts, post-trial, have
clearly announced: many of those babies should never have been admitted
to that hospital. They needed Level 3 care such as is only available at
a large teaching hospital.

The basic research unit in this case should have been the pregnancies,
the mothers; and the mother's antenatal care is extraordinarily
important. There were no obstetricians among the prosecution experts and
the mothers' obstetric case notes were not part of the medical evidence
brought into the trial.

This brings us to another statistical aspect: the spike in deaths. It is
very likely that closure of level 3 maternity units in North Wales in
Spring 2015 triggered a huge influx of high acuity patients into CoCH --
the neighbouring Level 3 units in Liverpool could not cope with the
increased demand. The spike in deaths at CoCH was statistically
significant, relative to that hospital. On the other hand, statistical
data shows that such large spikes in maternity units in the UK are
common; every year there is such an occurrence. It is cause for
investigation at the hospital and presumably, time and time again, an
explanation is found, not associated with criminal behaviour by a nurse.
Never done at the Countess of Chester was to compute the expected number
of deaths given the acuity of the babies admitted during the course of
the year during which Letby was allegedly on a killing spree. This could
be done; it could be done given the data which is routinely collected at
national level in the UK. It should have been done by the RCPCH, but the
team of medical experts which visited CoCH did not include a
statistician or epidemiologist.

\begin{figure}[t]\centering\includegraphics[width=\linewidth,height=0.4\textheight,keepaspectratio]{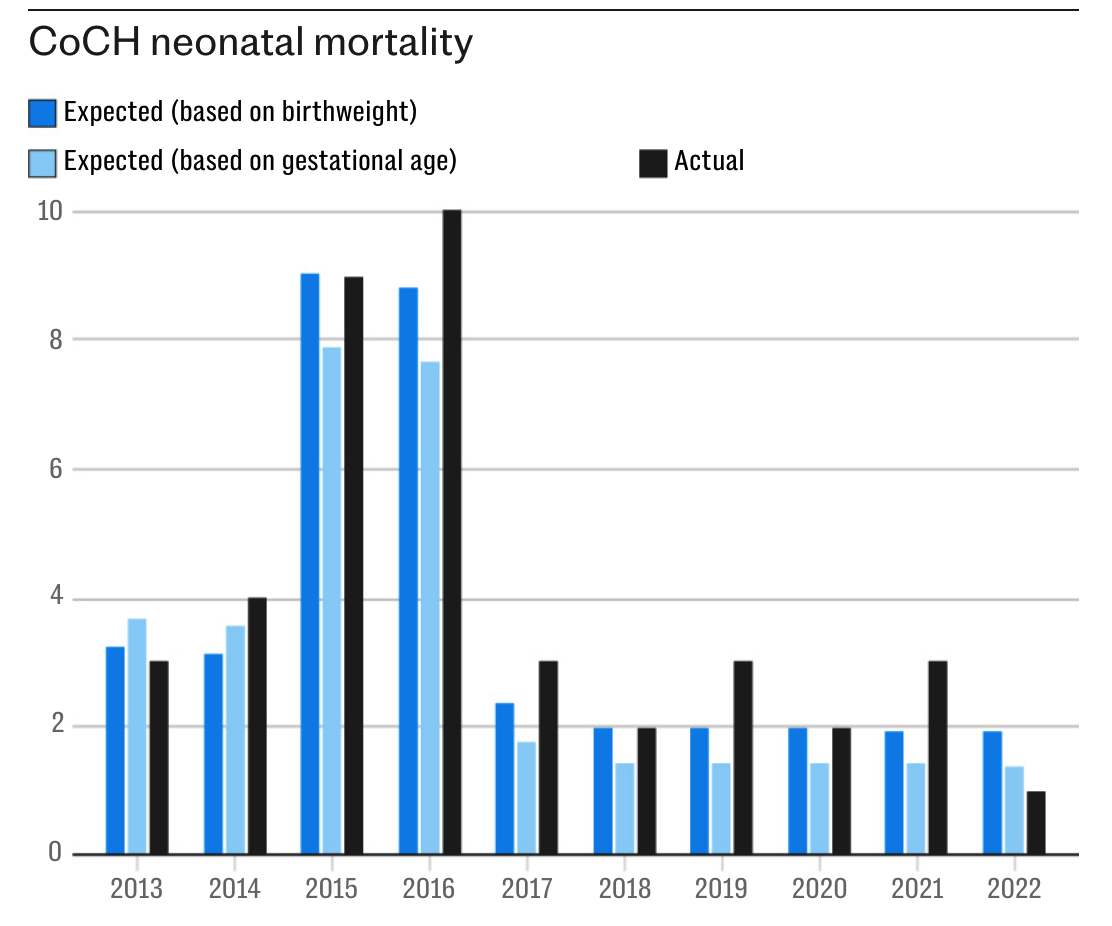}\caption{Small increases in acuity lead to a large increase in expected
mortality. (c) Sarah Knapton and Peter Elston. The Telegraph, 6
September 2024.}\end{figure}

The closest we have to a statistical analysis of the impact of acuity on
mortality at CoCH was done by statistician Peter Elston and journalist
Sarah Knapton in two newspaper articles, \citet{knapton2024}.
There exists publicly available aggregate data for all hospitals, and we
have aggregate data for the Countess of Chester. Unfortunately, publicly
available summary statistics are tabulated one variable at a time: we do
not have higher dimensional tabulations. For instance, we know how
mortality rata depends on birthweight and how it depends on prematurity
but not how it depends on both at the same time. The two variables are
highly correlated. The joint effect of a low gestation and birthweight
low for that gestation, is probably catastrophic. The condition is known
as Intrauterine Growth Restriction~(IUGR) and several of the babies in
the case suffered from it.

\citet{knapton2024} show the effect of a small increase in acuity
due to birth weight decrease, and similarly, due to gestation decrease.
Other statistical analyses are possible and should, in the opinion of
the author, be performed, in collaboration between medical experts and a
statistician. For instance, thanks to the Thirlwall Inquiry we now have
a list not just of the babies in the indictment, but also of 9 (!) other
babies who died in the same period, with summary information about the
cause. These further cases give the lie to the claim that the unit was
only treating babies with a good prognosis, in stable condition and
slowly improving. A small team of medical experts could, for each of the
babies about whom we have medical information concerning their state at
admission to the NICU, come up with a subjective determination of the
chance, given that information, that such a baby would not survive to
full term in an average UK Level 2 NICU. The sum of those chances is an
estimate of the number of deaths which should and could have been
expected among this subgroup of patients. The square root of the sum of
chance of death multiplied by its complementary chance of survival is an
estimate of the uncertainty in the first estimate of total mortality (in
the given subgroup of patients).

Hospital consultants told police investigators and later the court that
there was no increase in acuity of the babies being cared for on the
unit. Nurses on the other hand did report their own impression, which
was the opposite. More seriously ill patients mean longer stays in the
unit and higher occupancy, if the overall admission rate has not
changed. It also means higher hospital funding from the NHS due to NHS
accountancy practices. There are several motivations for a hospital unit
to take on more challenging patients: increase income, gain experience,
motivate an upward level classification.

The impact of an extraordinary probability computed by an apparently
professional statistician is illustrated by two other cases, both of
which were miscarriages of justice. In the case of the Dutch nurse Lucia
de Berk, everyone was impressed that in one year there had been 9
suspicious deaths and collapses, every one of which occurred in the
shifts of Lucia. Statistician Henk Elffers (at the time, with a PhD in
geography, not yet a professor) combined the data from the ward where
Lucia had worked most recently, with smaller data sets from two wards at
a previous hospital and announced that the probability of chance was 1
in 342 million. That number had enormous impact on everyone who heard
it. Medical experts later told the court that normally they would have
believed that certain incidents were natural, but because Lucia was
present, they decided that they were unnatural. The statistical evidence
had contaminated medical expert opinion.

Later it transpired that the statistical analysis was wrong on three
counts: events had been omitted which were not suspicious because Lucia
was not present. The method of combination of \textit{p}-values from three
2x2 contingency tables was incorrect and inflated the number by a factor
1,000. Confounding variables had not been taken account of. Taking
account of all three problems reduced 1 in 342 million to a not terribly
impressive 1 in 50.

In the case of Daniela Poggiali, a professor of epidemiology worked
closely with a forensic toxicologist. The epidemiologist presented
numbers and graphics which damned Ms Poggiali and the numbers were
widely publicised. Poggiali was perhaps responsible for 100 excess
deaths. The \textit{p}-value, measuring ``statistical significance'' was
again 1 in an astronomically large number.

Later it transpired that a large part of the association between deaths
and Poggiali's presence was completely spurious. Death certificates
reported the time at which an authorised person determines that a
patient has died. On a large ward with mainly terminally ill patients,
and several deaths a week, it turned out that many times of deaths were
actually the time, during the handover between two shifts, when an
authorised person checked on all the patients. Another such check was
performed at midnight, since the day on which someone dies has large
financial and legal consequences. Poggiali worked long shifts. She
typically clocked in for a new shift shortly before the handover at the
beginning and typically clocked-out shortly after the handover at the
end.

Fortunately, it was possible to illustrate these phenomena with simple
statistical graphics, and this led in part to Poggiali's exoneration.

In many cases the spike in mortality comes first, the association with a
particular nurse (by hospital consultants) is only made later. In the
Letby case, it seems that the initial deaths in June 2015 surprised and
shocked hospital consultants and the link with Letby was made by them
very soon after the first distressing and unexpected deaths.

\section{Forensic evidence}

In a number of these cases there is medical evidence suggesting
inflicted harm was done to one or two patients. We will discuss a couple
of examples, emphasising that the medical significance of tests done in
a hospital lab depends on statistical data and statistical analysis.

In the Allitt case and in the Letby case there is evidence of insulin
poisoning. The evidence is the result of hospital lab tests by the
so-called immunoassay method, routinely performed on a small blood
sample in cases where doctors are combating hypoglycaemia. Leading
forensic scientists have declared in major publications that the
immunoassay test is too unreliable for forensic purposes. It responds
not just to insulin but also to precursors of insulin, and to ``bound
insulin'', which is insulin which is no longer involved in glucose
metabolism but is in a sense neutralised by being bound to various
possible different antibodies in the blood. According to the literature,
one must painstakingly exclude all the known possible causes of a false
positive result. In case of suspected factitious insulin poisoning, it
is necessary to test the blood sample using the gold standard method,
liquid chromatography followed by tandem mass spectrometry. The local
lab which did the immunoassay tests in the Letby case telephoned the
hospital to communicate this information, as they are formally obliged
to do so according to official protocol. The doctors who asked for the
immunoassay tests did not proceed with a new test because the babies'
hypoglycaemia had already abated, and the babies soon after left CoCH.

In the Allitt case, the Guildford lab of Prof Vincent Marks verified the
immunoassay results by performing the necessary supplementary tests to
rule out interference by then known antibodies on further available
blood samples.

In the case of Lucia de Berk, the ``smoking gun'' forensic evidence was
a sample of watery blood obtained after postmortem analysis from the
body of the last patient in the case -- ``Baby Amber''. Her unexpected
death was the trigger which sparked hospital and then police
investigation. One of Lucia's fellow nurses told matron that she was
disturbed by the event, together with disturbing anecdotes which Lucia
had told her about her past life. The Netherlands forensic institute
analysed the sample and found the heart medication digoxin present, in
concentrations which suggested to a prominent toxicologist that Amber
had died of digoxin poisoning. Lucia was the last person to see the baby
alive, and it appeared that monitoring equipment had been switched off
by her long enough to enable her to give the baby digoxin.

Later, this was all disputed by several other toxicologists. The baby
had earlier been treated with dioxin which can be stored in various
organs in the body and released after death. Moreover, it was also
discovered that a further monitor had not been switched off and the
whole timing of the attack hypothesised by the prosecution was
overthrown.

The original toxicologist whose evidence sealed Lucia's fate, later
declared that he agreed with the new toxicologists. He did not retract
his diagnosis but asserted that the new experts had access to
information which had been withheld from him.

In the case of Daniela Poggiali, decisive evidence was a postmortem
determination of the concentration of Potassium Chloride in the aqueous
fluid (the fluid which fills the eyes), some hours after the death of
one of the patients. The last person who had attended to that patient
was Daniela Poggiali, who was known to have had quite some dislike for
her (probably mutual). The prosecution expert toxicologist had used data
collected from past cases to compute the rate at which the concentration
of potassium ions in the aqueous fluid slowly increases after death
towards the normally much higher concentration in the rest of the body.
He hereby could deduce the concentration at death and found such a high
concentration that he could definitively state that death was due to
potassium chloride poisoning.

\citet{dotto2021} were alerted to the
statistical aspect of this calculation by one of the medical experts
advising the defence. In fact, the computation depended on fitting a
curve to statistical data, and it had been performed by the same
epidemiologist who analysed the roster data. The duo (toxicologist and
epidemiologist) were long time scientific collaborators. Gill and
Mortera discovered that the statistical uncertainty in the calculation
was so large that it could not be excluded that the potassium
concentration at death was completely normal.

The moral of these examples is that the deductions of experts in
medicine and in toxicology depend on statistical data and can be subject
to large statistical uncertainty for many reasons, reasons which those
experts might be hardly aware of. A scientific expert is obliged to
report uncertainties and alternative explanations to a court. Their duty
is to the court, not to the party which instructed them. Many experts
however are naturally eager to please the party who is paying them, and
moreover eager to present much higher confidence in their findings than
is justified. A highly confident expert makes a much more powerful
impression on a lay jury than an expert who is honest and open about
uncertainties. Police investigators and prosecutors learn from
experience which experts are likely to best help them gain a
prosecution.

\section{Character evidence}

A killer nurse evidently must be an evil person, so evidence is sought
by police investigators and later brought into the trial for
psychopathy.

In the case of Lucia de Berk, diary writings of Lucia in which she spoke
mysteriously of a secret compulsion, which must never be revealed to
anyone, but which should go to the grave with her, were interpreted as
meaning that she had a compulsion to kill \citep{gill2018}.

Her explanation was that she had a compulsion to offer patients (elderly
or otherwise terminally ill persons at the end of their lives) a tarot
card reading. She owned a set of tarot cards and was fascinated by
tarot, but she knew that for a nurse this activity was unprofessional.
Her explanation was confirmed by a court appointed psychiatrist as
completely plausible.

In the Lucy Letby case, scribblings on post-it notes which we now know
were part of a therapeutical exercise advised by a HR staff member
contained sinister remarks to the effect ``I am evil, I did this'' but
also ``I'm innocent'', ``why am I being persecuted'', and so on. The
notes are crammed full of phrases organised in columns, probably
separating her thoughts concerning her innocence and suffering from what
other people were saying about her; thoughts which filled her mind to
distraction. A psychiatrist expert in false convictions has stated that
the notes are of no evidential value whatsoever, and that Letby's own
explanation of them is completely natural.

Beverley Allitt's adolescence was troubled. She practiced self-harm.
Much evidence was presented at her trial of disturbed behaviour. It has
been suggested that she was chosen as a scapegoat because of her
psychological vulnerability. The present author, as a statistician, is
not qualified to enter that debate. He does presently suspect she is innocent too,
and recommends the reader consult the recent book \citet{ross2025}.

\section{Conclusion}

UK convicted nurses Victorino Chua, Ben Geen, and Colin Norris were
previously mentioned but not used as examples. In fact, their cases all
contain many very disturbing features, and furnish further examples of
the three features discussed above. First and foremost, the evidence
against them is largely statistical. There is some medical evidence that
some patients may have been deliberately harmed, but it is not
conclusive. The hypothesis that the convicted nurse himself inflicted
that harm is just that: a hypothesis. Character evidence played a huge
role in these nurse's convictions but does not appear to this author, a
mathematician and a scientist, to be much more than storytelling after
the fact. Post conviction, the defence story is slowly erased from go-to
sources like Wikipedia; and the prosecution story is embellished by
``true crime'' writers, in books and in documentaries.

In conclusion, we cannot do better than to summarise RSS
recommendations. In suspected medical misconduct~cases, investigations
should be supervised by multidisciplinary expert panels independent of
both the suspect and their employer and including professional
statisticians or epidemiologists. Confirmation bias and the Texas
sharpshooter paradox must be avoided through use of rigorous forensic
investigation protocols, in which the selection of ``suspicious cases''
must be fully reproducible through explicit and objective criteria, and
reproducible and exhaustive search procedures. Interaction between the
legal and statistical communities must be fostered by the leaders of
their respective communities.

\clearpage

\nocite{*}
\bibliography{document}
\end{document}